# High Energy Vision: Processing X-rays

Joseph DePasquale[1], Kimberly Arcand[1], Peter Edmonds[1]

[1]Smithsonian Astrophysical Observatory, 60 Garden St, Cambridge, MA 02138, USA

Correspondence: Joseph DePasquale, Smithsonian Astrophysical Observatory, 60 Garden St, Cambridge, MA 02138, USA. Tel: 617-496-7912.

## Abstract

Astronomy is by nature a visual science. The high quality imagery produced by the world's observatories can be a key to effectively engaging with the public and helping to inspire the next generation of scientists. Creating compelling astronomical imagery can, however, be particularly challenging in the non-optical wavelength regimes. In the case of X-ray astronomy, where the amount of light available to create an image is severely limited, it is necessary to employ sophisticated image processing algorithms to translate light beyond human vision into imagery that is aesthetically pleasing while still being scientifically accurate. This paper provides a brief overview of the history of X-ray astronomy leading to the deployment of NASA's Chandra X-ray Observatory, followed by an examination of the specific challenges posed by processing X-ray imagery. The authors then explore image processing techniques used to mitigate such processing challenges in order to create effective public imagery for X-ray astronomy. A follow-up paper to this one will take a more in-depth look at the specific techniques and algorithms used to produce press-quality imagery.

**Keywords**: image processing, data, X-ray astronomy, science communication

## 1. A brief history of X-ray Astronomy

In its relatively short history spanning less than 70 years, the field of X-ray astronomy has seen a dizzying rate of technological advancement, delivering spectacular results across a wide range of areas. The visual communication of these results is a key motivation for the work described in this paper.

High-energy radiation in X-rays and gamma-rays are absorbed by Earth's atmosphere (figure 1), so that X-ray astronomy depended on advancements in rocketry and space science to be fully explored (Tucker & Tucker, 2001).

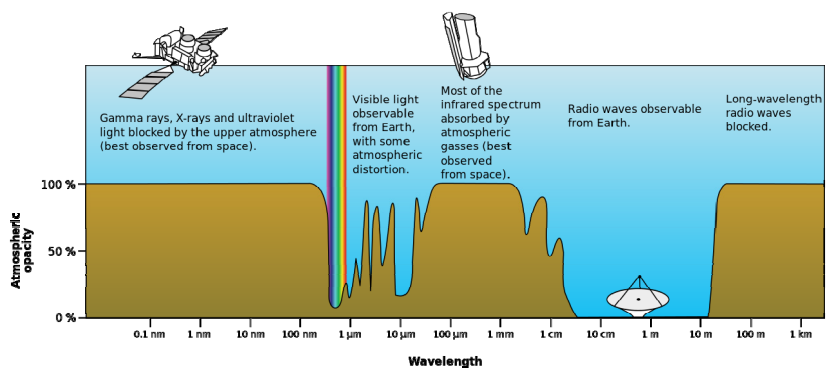

Figure 1. Atmospheric absorption across the electromagnetic spectrum. High energy waves past the visible spectrum are absorbed high above Earth's surface, hence the need for space-based observatories at these wavelengths. Image Credit: NASA (public domain)

About 16 years after the first experiments establishing sources of extra-solar X-rays (Giacconi, Gursky, Paolini, & Rossi, 1962), NASA launched a series of scientific instruments into Earth orbit, including the first fully imaging, space-based X-ray observatory named *Einstein*. With an angular resolution of 3 to 5 arcseconds, *Einstein* resolved and accurately located over 7000 X-ray sources, including stellar coronas, X-ray binaries, galaxies and quasars, and showed that most of the X-ray background is due to discrete sources (Gorenstein, Harnden, & Fabricant, 1981).

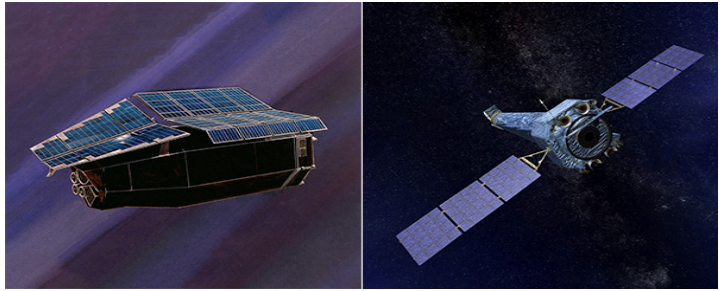

Figure 2. Artist's conceptions of the *Einstein* Observatory (left) and the Chandra X-ray Observatory (right). Image Credit: (left) Einstein: NASA (public domain); (right) Chandra: NGST

Over the course of the next 20 years, development of the Advanced X-ray Astrophysics Facility (AXAF), which would eventually be renamed as NASA's Chandra X-ray Observatory, moved at a steady pace (Weisskopf, ODell, Elsner, & Speybroeck, 1995) & (Canizares, 1990). In the 37 years between the first detection of extra-solar X-rays, and the launch of Chandra in 1999, the sensitivity of X-ray detectors had increased over 10 billion times.

Chandra's unprecedented angular resolution of 0.5 arcseconds coupled with increased detector sensitivity has helped transform our understanding of the high-energy Universe in the observatory's 15 years and counting in orbit (Weisskopf et al., 2003, Swartz, Wolk, & Fruscione, 2010 & Tananbaum, Weisskopf, Tucker, Wilkes, & Edmonds, 2014). Chandra is capable of making detailed X-ray images of star clusters, supernova remnants, quasars, and collisions between clusters of galaxies. Chandra has probed the geometry of space-time around black holes (Miller et al., 2002), traced the dispersal of iron and other elements by supernovas (Hwang & Laming, 2012), revealed that whirling neutron stars only twelve miles in diameter can generate streams of high-energy particles that extend for light years (Hester et al., 2002), and explored the interactions between exoplanets and their host stars (Schröter et al., 2011 & Poppenhaeger, Schmitt, & Wolk, 2013). On a larger scale, Chandra has helped to confirm that galaxy clusters and the Universe are dominated by dark matter and dark energy (Allen, Schmidt, Ebeling, Fabian, & Speybroeck, 2004, Clowe et al., 2006 & Vikhlinin et al., 2009). Chandra is at the vanguard of our current understanding of the mysterious nature of dark matter and dark energy (Bulbul et al., 2014 & Boyarsky, Ruchayskiy, Iakubovskiy, & Franse, 2014).

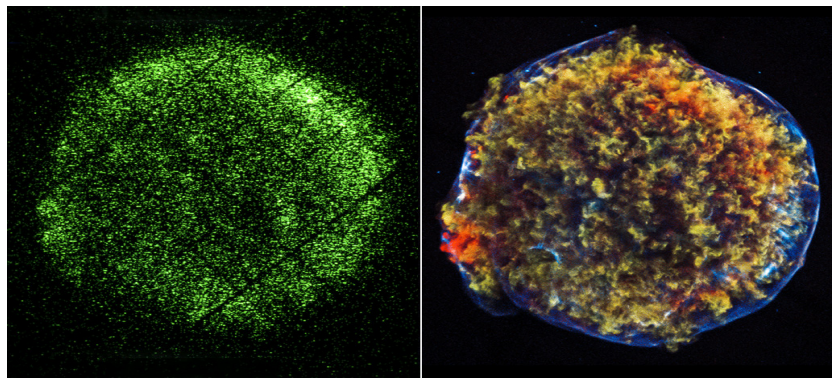

Figure 3. Comparison of Einstein's view of Tycho's Supernova Remnant (left) to that of Chandra's (right) illustrating the dramatic advancements in X-ray astronomy in the era of Chandra. Image Credit: (left) NASA/CfA; (right) NASA/CXC/SAO

## 2. X-ray Image Processing Overview

Despite the impressive achievements described in the previous section, creating compelling imagery for the greater public from X-ray data poses a unique set of challenges. There is an immediate hurdle of explaining how the observatory can "see" in X-rays. Representing the sometimes amorphous objects in X-rays in ways that the non-expert can identify with is another challenge for Chandra's office of public engagement. Much care is taken to represent these objects truthfully, and translate the data from something the human eye was never meant to sense, into an accessible and aesthetic view of the Universe (L. F. Smith et al., 2010 & Arcand et al., 2013).

In the following sections we provide an overview of the methods that have been developed and fine-tuned by Chandra's Communications and Public Engagement group to produce X-ray images for non-experts that attempt to maximize aesthetic beauty while telling the story of the science behind the data.

*2.1 Photon Starvation*

X-ray astronomy has the peculiar distinction of being starved of photons; the sky is simply not very bright in X-ray light (Schwartz, 2009). X-ray observations lasting more than 5,000 seconds can routinely end up yielding less than a thousand individual X-ray photons for a source, depending on its intensity. Because the distribution of X-ray photons is Poisson in nature and a source signal can be very weak, a great deal of background can be a typical signature of an X-ray observation of extended sources with lower surface brightness (Siemiginowska, 2009). Balancing noise-reducing smoothing techniques with the inherent resolution of the detectors is of prime concern when dealing with noise as preserving the highest resolution possible across the image will produce a higher quality final result. Described as photogenic resolution in Christensen, Hainaut, & Pierce-Price (2014), the size of finest features that can be resolved across the field of view of an image is one of the key factors in determining the aesthetic quality of that image. With this in mind, we explore the first stages of image preparation after pulling data from the archive server.

*2.2 Dealing with Raw Data*

The Chandra data archive plays a central role in the storage and distribution of Chandra data to all users (McCollough, Rots, & Winkelman, 2006). The archive is open to the public and most data are freely available after a one year proprietary period. Downloading data from the archive is a simple procedure of providing a query and deciding which datasets are needed. From there, data products can be downloaded and are then ready for further processing. The native format of Chandra data is a collection of Flexible Image Transport System (FITS) formatted files (Pence, Chiappetti, Page, Shaw, & Stobie, 2010).

*2.3 Chandra Interactive Analysis of Observations (CIAO)*

CIAO is a comprehensive software suite designed with processing Chandra data in mind (Fruscione et al., 2006). CIAO is also flexible and can be applied to other uses as well. CIAO is what most observers use when preparing data for analysis and interpretation. We use CIAO tools in the first stages of preparing Chandra imagery for press releases and other public uses.

*2.4 Merging Observations*

The typical public image will combine data from several Chandra observations. CIAO provides the tools needed to merge multiple datasets into a single image maximizing the signal to noise ratio for a given observation. A Chandra image is a FITS data file named the events list. It is, essentially, a comprehensive list of all X-ray photons collected during a given observing period. For each photon collected, the events list contains an entry detailing the exact time, location on the detector, and energy of that particular photon. Compiling a two-dimensional detector view of all photons in an observation yields the X-ray image of that source.

*2.5 Creating a Flux Image*

The flux image is the springboard for the creation of the public image. Once the data have been merged, and the events list has been converted into a two-dimensional image, there may be instrumental artifacts present in the image. These can be mitigated by creating an energy dependent "exposure map" to account for artifacts like gaps in the CCDs, bad columns, vignetting of the detector effective area, etc. The flux image is the cleanest possible form the image can take in its raw state. Even so, background noise and other instrumental artifacts may still be present in these images (figure 4) and now the challenge begins to maximize aesthetic appeal while maintaining a truthful representation of the science data. Depending on the nature of the source data, either a single-color broadband image will be created, or several images in different energy bands can be created to ultimately yield a full color image.

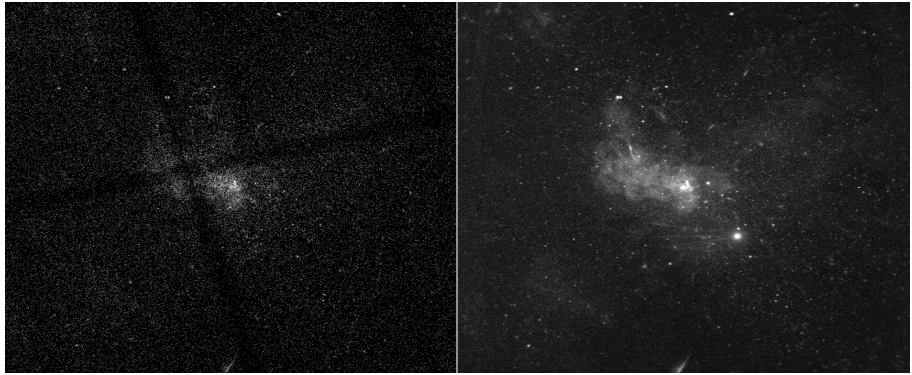

Figure 4. A single observation (14 hours integration time) of Sagittarius A* is compared with a merged image combining data from 43 observations (278 hours integration time). Note the improved signal to noise of the image on the right, but also the presence of "readout streak" artifacts on the bright point source to the lower right of center even after initial processing. Image Credit: NASA/CXC/SAO

*2.6 Intermediate Processing*

Once data preparation in CIAO has been completed, we move the image(s) into a digital image processing package using several software solutions including Pleiades Astrophoto's PixInsight, and Adobe Photoshop[1]. The task of generating an image for release can be distilled into the following steps (Rector, Levay, Frattare, English, & Puuohau-Pummill, 2007):

1. intensity scaling and projection of the data into a gray-scale image

2. importing as layers into an image processing software package

3. intensity rescaling layers to increase contrast

4. assigning color to each layer

5. fine tuning the image, including color balance, removal of artifacts, orientation and framing

6. preparing the image for electronic distribution and print production.

*2.7 Image Smoothing & Intensity Scaling*

Mitigating the noise in X-ray imagery can be done through the careful use of context-sensitive smoothing algorithms. These algorithms preserve bright point sources and contours in an image while smoothing out the background noise. Particular attention must be paid when applying smoothing algorithms such as adaptive kernel smoothing. If too much smoothing is applied, or the parameters used do not fit the source data, smoothing artifacts may be introduced (Ebeling, White, & Rangarajan, 2006). Potential artifacts can be seen in figure 5 on the top right, where low surface brightness features can fool the smoothing algorithm into creating structures which do not actually exist.

---

[1] The Chandra X-ray Center and the Harvard-Smithsonian Center for Astrophysics do not endorse any particular vendor's products. Please read all license agreements before downloading any software.

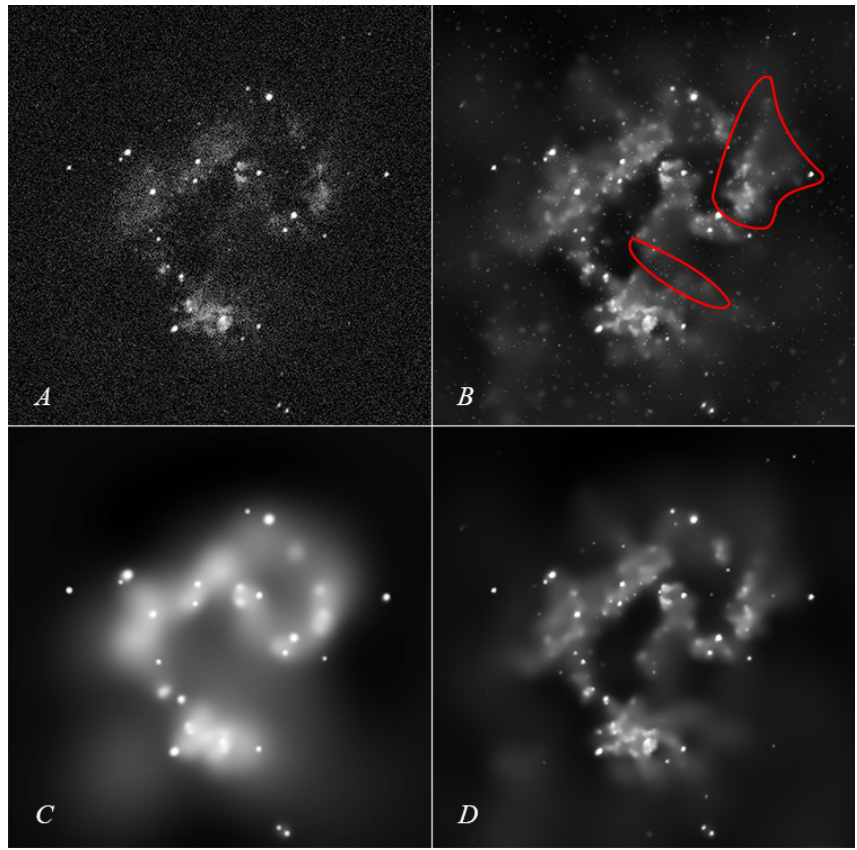

Figure 5. A comparison showing the Antennae Galaxies (NGC 4038/4039) in X-rays as seen by Chandra. Panel A shows the raw image with no smoothing applied. In panels B, C, and D, an adaptive kernel smoothing algorithm (Ebeling, White, & Rangarajan 2006) has been applied which is designed to preserve brighter point sources and the structures around them while smoothing out the background noise and diffuse X-ray emission. In panel B the input signal to noise variable was chosen to be too low and the algorithm has created structural artifacts like those seen in the red regions (spurious point sources and faint "tubes" of emission). Panel C shows a version of the image that has been over-smoothed where the signal to noise used by the algorithm was too high. Panel D shows the amount of smoothing that was used for the press image. Image Credit: NASA/CXC/SAO

Due to the extreme dynamic range in astronomical images, a scaling algorithm needs to be applied to balance the ratio of the brightest to faintest features in the image and allow the full dynamic range of the pixels to be visible (Hurt & Christensen, 2007). A poorly scaled image will introduce artifacts in the form of clipping the brightest parts of the image to pure white and losing detail and definition in the final image.

Proper smoothing and scaling of the raw data is an iterative process where the image processor must make decisions as to how best represent the data without introducing new artifacts.

### 3. Generating Publication Quality Imagery

*3.1 A Spectrum of Color*

Color can be added to astronomical imagery in a process known as representative coloring (also called false coloring). The addition of color to a greyscale image not only heightens the aesthetic appeal of the image, it can also serve as an important dimension of data visualization, adding value and meaning for the viewer (L. Smith, Arcand, Smith, Smith, & Bookbinder, 2015). Many Chandra images use color to denote intensity of X-ray light at different energies (e.g., where red represents soft X-rays, green represents medium energy X-rays and blue represents high energy X-rays - see figure 6). Using color in this manner not only tends to create a more attention-getting and/or pleasing image, it also reveals important information about the spatial distribution of X-ray energies in a particular data set. Other times, color is used to denote the presence of certain heavy elements, or specific topographical features of a data set. The authors regularly make use of insights gained from studies of the perception of color in astronomical imagery to inform decisions regarding color choice in Chandra data (Arcand et al., 2013).

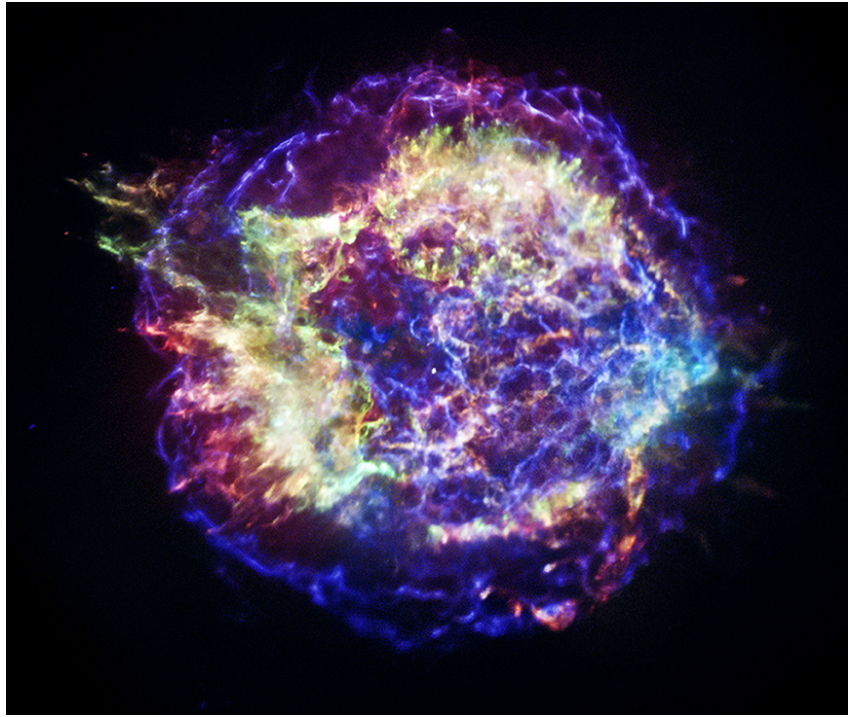

Figure 6. Chandra's megasecond view of Cassiopeia A. In this color image, soft energy X-rays from 0.5-1.5 keV are colored red, medium energy X-rays from 1.5-2.5 keV are green, and high-energy X-rays from 4.0-6.0 keV are colored blue. Image Credit: NASA/CXC/SAO

*3.2 Multi wavelength Imagery*

In addition to adding color to images, leveraging the power of multiple wavelengths can add depth and dimension, as well as context, to astronomical imagery. This is particularly effective in the X-ray regime as additional wavelengths can enhance the photogenic resolution of astronomical imagery as well as add visual context for non-experts. Modern astronomers have access to terabytes (Brunner, Djorgovski, Prince, & Szalay, 2001) of information with unprecedented, high-quality views of our cosmos from observatories spanning the entire electromagnetic spectrum that make it possible to achieve a holistic view of physical processes and morphological structures.

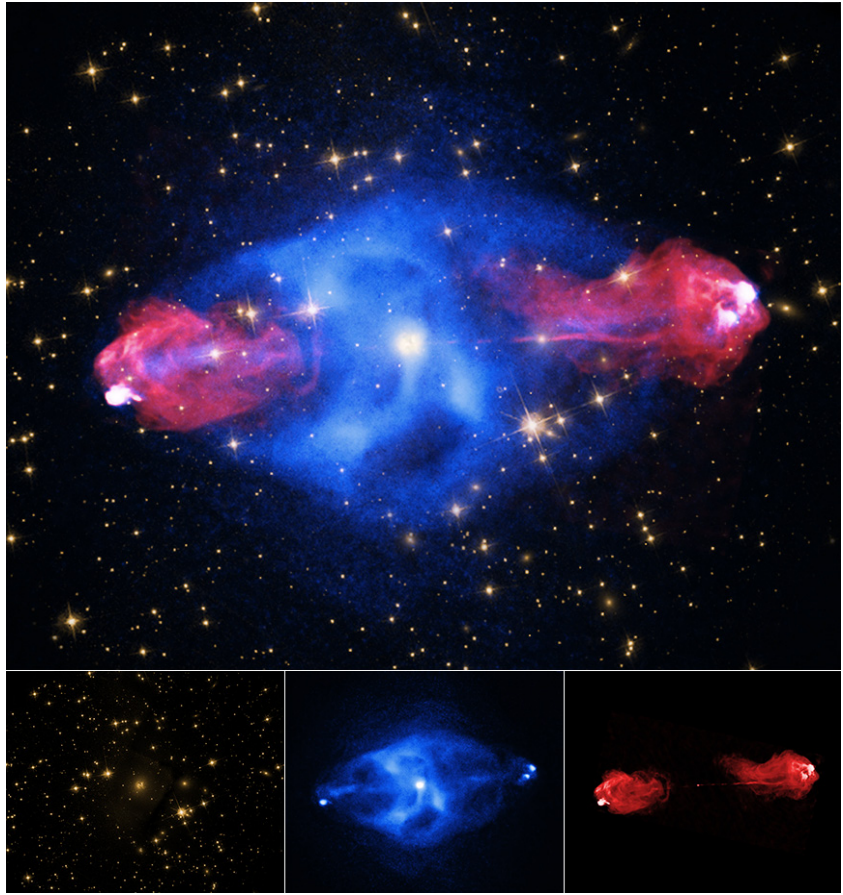

Figure 7. A multi-wavelength view of the galaxy in Cygnus A. In the top, composite image, optical data from the Hubble Space Telescope are represented by gold, while X-ray (Chandra) and radio (VLA) data are shown in blue and red respectively. The bottom three panels show the individual images of the composite. Image Credit: X-ray: NASA/CXC/SAO; Optical: NASA/STScI; Radio: NSF/NRAO/AUI/VLA

The complementary structures seen in figure 7 add a new dimension to the X-ray image. Confined to optical wavelengths, the jet structure of this galaxy is hidden. The jets appearing in the X-ray and radio images drive the morphology of the large cavities seen in the X-ray image. This is a clear example of how multi-wavelength astronomy illuminates the story behind the science of the image and, in nearly all cases, merging multiple wavelengths into a composite image simultaneously serves to enhance the aesthetic appeal of the image.

*3.3 Fine Tuning and Final Preparation*

At this point in the process, fine-tuning of the color balance and contrast as well as removing any last remnants of instrumental artifacts is necessary before the image can be considered ready for the public (figure 8). Ensuring that there are no systematic color biases in the background regions of the image will help to maintain a properly color balanced image and can be achieved by adding or removing color as needed. For example, if the image appears too blue, the complementary color of yellow can be added to achieve color balance. Additionally, overall contrast adjustments at this stage will preserve a compelling and eye-catching dynamic range. The final step in the production process is to produce a flattened image in several print and web-ready formats suitable for publication (Rector et al., 2007).

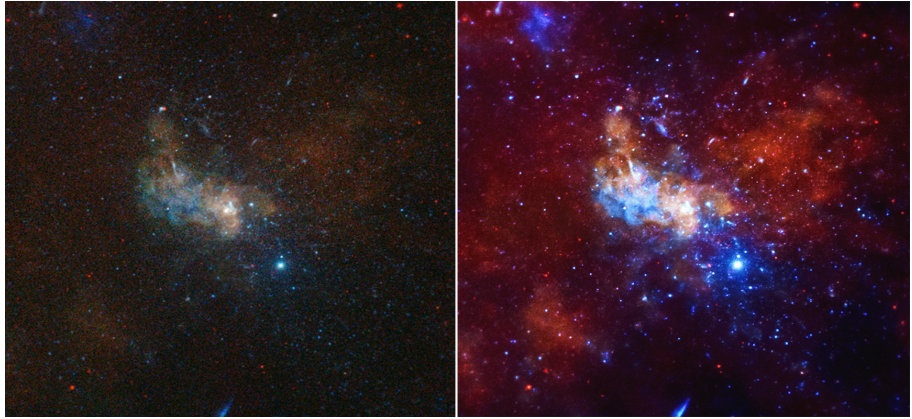

Figure 8. A before and after comparison showing the same field of view of Sagittarius A* with the initial three-color combination on the left compared to the final public image after image smoothing, intensity, contrast and color adjustments. Both images are color coded by X-ray energies: Red (2-3.3 keV), Green (3.3-4.7 keV), Blue (4.7-8 keV).
Image Credit: NASA/CXC/SAO

**4. Summary**

The field of X-ray astronomy has seen dramatic growth over the past 70 years inspiring the technological advancements that made NASA's Chandra X-ray Observatory possible. Chandra's unprecedented imaging capabilities have provided an extraordinary window into the high-energy Universe. Simultaneous advancements in computing power and image processing algorithms provide the additional tools needed to overcome X-ray image processing hurdles to produce attractive images of the Universe from Chandra data as well as from data of other world-class observatories. Such public images can serve an integral part of the legacy of these observatories by visually telling the stories of complex scientific achievements as well as nuanced interim advancements. The responsibility lies with the image processing team to balance the beauty of the data with the scientific story behind them.

**Acknowledgements**

The Chandra X-ray Center is operated for NASA by the Smithsonian Astrophysical Observatory. This paper was developed with funding from NASA under contract NAS8-03060.